# Unusual magneto-transport of $YBa_2Cu_3O_{7-\delta}$ films due to the interplay of anisotropy, random disorder and nanoscale periodic pinning

J. Trastoy[1], V. Rouco[1,2], C. Ulysse[3], R. Bernard[1], A. Palau[2], T. Puig[2], G. Faini[3], J. Lesueur[4], J. Briatico[1] and J.E. Villegas[1,†]

[1]*Unité Mixte de Physique CNRS/Thales, 1 avenue A. Fresnel, 91767 Palaiseau, and Université Paris Sud, 91405 Orsay, France.*

[2]*Institut de Ciència de Materials de Barcelona, ICMAB-CSIC, Campus de la UAB, E-08193 Bellaterra, Spain.*

[3]*CNRS, Phynano Team, Laboratoire de Photonique et de Nanostructures, route de Nozay, 91460 Marcoussis, France.*

[4]*LPEM, CNRS-ESPCI, 10 rue Vauquelin 75231 Paris, France.*

We study the general problem of a manifold of interacting elastic lines whose spatial correlations are strongly affected by the competition between random and ordered pinning. This is done through magneto-transport experiments with $YBa_2Cu_3O_{7-\delta}$ thin films that contain a periodic vortex pinning array created via masked ion irradiation, in addition to the native random pinning. The strong field-matching effects we observe suggest the prevalence of periodic pinning, and indicate that at the matching field each vortex line is bound to an artificial pinning site. However, the vortex-glass transition dimensionality −quasi-2D instead of the usual 3D− evidences reduced vortex-glass correlations along the vortex line. This is also supported by an unusual angular dependence of the magneto-resistance, which greatly differs from that of Bose-glass systems. A quantitative analysis of the angular magneto-resistance allows us to link this behaviour to the enhancement of the system anisotropy, a collateral effect of the ion irradiation.

†email: javier.villegas@thalesgroup.com

1. **Introduction**

A number of fundamentally and technologically relevant problems in Physics can be mapped to a manifold of elastic, three dimensional interacting elements (e.g. "strings" or "walls"), whose ordering and dynamics depend on the presence of pinning, thermal activation, and some form of line tension that opposes to their deformation. Colloids [1], flux lines in type-II superconductors, liquid/solid interfaces [2], domain walls and skyrmions in ferroic materials (ferroelectrics [3] and ferromagnets [4-6]) are just a few examples. A problem common to all of them is how the presence of pinning disrupts the natural ordering and provokes the deformation of the individual elements –thereby affecting their spatial correlations– and how this changes the manifold's phase diagram. In this paper we address a problem along these lines, using as a model vortices in a high critical temperature ($T_c$) superconductor with artificial pinning.

The vortex phase diagram in the mixed state of superconductors is determined by the balance between elastic energy (inter-vortex interactions and vortex line tension), thermal energy and pinning energy. The strong thermal fluctuations and anisotropy characteristic of high-$T_c$ superconductors yield a rich vortex phase diagram [7], on which pinning effects have been extensively studied for over two decades now [8]. Many different types of pinning defects have been considered, both natural and artificially-induced, which can be sorted into two groups: point and extended defects. Among the latter, the so-called correlated defects consist of distributions of equally oriented anisotropic structures, each of which can entirely accommodate a vortex line when the magnetic field is applied parallel (or close) to a particular direction [9]. The tracks created in cuprate superconductors via heavy-ion irradiation [10-13] are a good example of these. In the vast majority of the existing studies, all sorts of pinning defects –correlated or not– are randomly distributed within the bulk of the material. Contrarily, and despite the interest raised by theoretical studies [14,15], the effects of spatially o*rdered* pinning on the vortex phase diagram of oxide superconductors have seldom been addressed experimentaly [16-18].

Despite the fact that high-$T_c$ superconductors present a more appealing vortex phase diagram than low-$T_c$ ones, most of the experimental work on *ordered* pinning has actually been made with the latter. For low-$T_c$ materials, a variety of nano-fabrication techniques allow easily creating ordered arrays of holes [19,20] or non-superconducting inclusions [21,22], whose sizes are comparable to the superconductor characteristic lengths. These arrays of extended defects produce very efficient vortex pinning [23], and have enabled the investigation of a number of phenomena [24], including commensurability [19-25], controlled vortex motion [26-28], switchable pinning [29] and vortex phase-transitions [30-32]. Regarding high-$T_c$ superconductors, in which nanofabrication techniques are more difficult to implement, attention has been paid to the critical current enhancement due to geometric matching of the flux-lattice to the artificial defect arrays [33-36], as well as to dynamic effects such as guided vortex motion [37] and rectification [38-40]. However, fewer studies have addressed how the presence of those arrays changes the vortex phase diagram [16-18]. All of them are based on the highly anisotropic $Bi_2Sr_2CaCu_2O_{8+\delta}$ (BSCCO), and exclusively address the situation in which the external magnetic field −and the induced vortices− are parallel to the c-axis (i.e. perpendicular to the defect array plane). Early experiments on single-crystalline BSCCO thin films with periodic arrays of through-holes [16] found that, unexpectedly, the matching effects between the flux-lattice and the hole arrays could be observed well above the melting line that separates vortex-solid and liquid phases, suggesting a form of pinned liquid. In experiments on BSCCO nano-ribbons with periodic through-hole arrays [17], strong matching effects were indeed observed at both sides of the melting line, near which they were significantly enhanced. Other experiments on thick BSCCO single-crystals with periodic arrays of surface holes [18] demonstrated deep changes of the vortex matter phase diagram and suggested the possibility of a Mott insulator phase when the number of flux quanta equals the number of pins. The Mott insulator phase is a particular case of the Bose-glass [9], in which all of the vortices are localised within *correlated* defects, and for which the phase transition with increasing temperature is into a delocalised state.

In this paper, we study $YBa_2Cu_3O_{7-\delta}$ (YBCO) thin films with artificial periodic pinning. Many aspects of the present study are in contrast with the experiments on BSCCO summarised above. First, pristine optimally doped YBCO presents a moderate anisotropy. For this reason, the c-axis vortex correlations are stronger, which confers vortex matter a three-dimensional character [41] that contrasts with the purely two-dimensional one of plain BSCCO [42]. Second, due to the presence of strong native *random* pinning, YBCO thin films typically show [43,44] a continuous vortex-glass transition [45] separating a liquid phase from a glass phase, instead of the first-order solid-to-liquid transition usually observed in BSCCO single-crystals [18] and in twin-free YBCO single crystals [46]. Third, in the present samples the artificial ordered pinning sites are created via masked ion irradiation [36], which allows us to obtain considerably denser pinning arrays than in earlier studies [16-18], and produces strong field matching effects within an unusually wide range of temperatures [36]. In the present paper, we investigate whether the masked ion irradiation modifies the glass transition typically observed in thin YBCO films −and in particular, whether a Bose-glass [9] rather than a vortex-glass [45] is stabilised. This is done via magneto-transport measurements in tilted applied fields, which we analyse using a scaling of V-I characteristics and a model we have developed to fit the resistance dependence on the applied field direction. Contrary to what one would have naively anticipated in view of the strong matching effects, we find the magneto-transport not supportive of a Bose-glass, but of a vortex-glass transition of quasi-two dimensional character. As detailed below, our analysis show that although periodic pinning enhances vortex ordering in two dimensions (in-plane), the concurrence of native *random* pinning and a relatively high anisotropy —that makes vortex lines "softer"— frustrates the correlations along the vortex line. This precludes the observation of the typical Bose-glass behaviour [8,10-12,47-51]. Our experiments illustrate the relevance of the balance between different pinning sources and the elastic properties, which may be extrapolated *mutatis mutandi* to other problems such as the smoothing of domain walls in ferromagnets via the introduction of artificial pinning [5].

## 2. Experimental

The periodic vortex pinning potentials are defined via a combination of e-beam lithography and $O^+$ ion irradiation in 50 nm thick c-axis YBCO films grown by pulsed laser deposition on (0 0 1) $SrTiO_3$ (STO). A detailed description of the fabrication method can be found elsewhere [36]. In summary, a mask is defined via e-beam lithography in a thick resist (PMMA) that covers the YBCO film. In the present experiments, the mask contains a square array of holes (diameter $\varnothing \sim 40$ nm) in which the inter-hole distance (centre to centre) is $d=120$ nm. A typical scanning electron microscopy image of one mask is shown in the inset of figure 1. Irradiation with $O^+$ ions (energy $E=110$ keV) through the mask induces disorder in the YBCO film, in particular oxygen vacancies and interstitials. Note that contrary to masked ion milling techniques [52], here the material is not etched and the sample surface is not affected [53] by the ion irradiation: only point defects are created in the bulk of the material. These defects are strongly concentrated in the regions directly exposed to the ion beam through the mask holes. Note that the 110 keV $O^+$ ions' track length into YBCO (~150 nm) largely exceeds the films thickness, which ensures that the point defects are created throughout the YBCO film, from its surface down to the STO substrate. By using a high enough irradiation fluence ($5 \cdot 10^{13}$ ions·cm$^{-2}$ in the present experiments) the superconducting critical temperature $T_c$ is locally depressed in the hole areas directly exposed to the ion beam, which makes of these areas strong extended pinning centres for flux quanta [36]. The latter is evidenced by mixed-state magneto-transport measurements (figure 1), which shows periodic field-matching effects. To perform those measurements, a multi-probe bridge (200 μm long and 40 μm wide) was lithographed and ion-etched in the samples. The magneto-transport experiments were carried out in two different cryostats: i) a He flow cryostat equipped with a 0.5 T electromagnet and a rotatable sample holder (±0.5 deg precision); and ii) a He bath cryostat equipped with a 7 T superconducting coil and a rotatable sample holder (±5 deg precision). Figure 1 shows a typical magneto-resistance curve at $T<T_c$, with the magnetic field applied perpendicular to the film plane (parallel to the YBCO c-axis). The curve shows a series of periodic oscillations with minima at the "matching fields" $B=\pm nB_\phi$, with $n$ an integer or a semi-integer, and $B_\phi=\phi_0/d^2=0.144$ T the field at which the density of flux quanta equals the density of holes in the square array. In the curve of figure 1, which corresponds to the sample studied in the present experiments, clear minima are observed for $n=\pm 1,2$, and weaker ones (as expected [20,54]) for $n=\pm 0.5$. At the matching fields, the commensurability between the flux lattice and the pinning array leads to enhanced vortex pinning [19-24,29-36,39]. Note that no periodic field-matching effects were reported by other groups that used ion irradiation to pattern high-temperature superconductors [52,55,56]. Throughout the paper, we will investigate how the commensurability effects observed here affect the glass transition and the angular dependence of the magneto-resistance.

### 3. Results

*3.1. Analysis of V-I characteristics.*

*V-I* characteristics were measured in several applied fields. For each particular field, a set of isotherms was measured within a temperature range. An example of the data sets typically obtained for each applied field is shown in figure 2, which corresponds to $B = 2B_\phi$. For the highest temperatures, a finite Ohmic resistance is observed in the low-current limit, as expected in the thermally activated flux-flow regime [41]. As the temperature is reduced, the Ohmic regime gradually disappears, and yields to a non-linear behaviour within the entire experimental window, and to vanishing resistance in the low-current limit, $\lim_{I \to 0} V/I = 0$. This is expected for a continuous (2$^{nd}$ order) vortex-glass transition [45], as well as for a Bose-glass transition [9].

We applied to each set of *V-I* curves the scaling analysis proposed in the vortex-glass [45] and Bose-glass theories [9]. For the vortex-glass, a data collapse of the *V-I* measurements should be observed according to the scaling ansatz [45]:

$$V\xi_{VG}^{2+z_{VG}-D}/I = \chi_{\pm} (I\xi_{VG}^{D-1}/T) \qquad (1)$$

where $\xi_{VG} \propto |1 - T/T_g|^{-\nu_{VG}}$ is the vortex-glass correlation length (which diverges at the glass transition temperature $T_g$), $T$ is the temperature, $\nu_{VG}$ and $z_{VG}$ are respectively the static and dynamic critical exponents, $D$ is the vortex-glass system's dimensionality (*D*=3 for a three-dimensional system), and $\chi_{\pm}$ are the scaling functions above (+) and below (-) $T_g$. For the Bose-glass, the scaling ansatz is [43]:

$$V\xi_{BG}^{2(z_{BG}-2)/3}/I = \chi_{\pm} (I\xi_{BG}^{2}/T) \qquad (2)$$

where $\xi_{BG} \propto |1 - T/T_g|^{-3\nu_{BG}/2}$ is the Bose-glass correlation length.

Note that if one is able to find a set of parameters $D$, $T_g$, $\nu_{VG}$ and $z_{VG}$ that allows scaling the data according to the vortex-glass model, the scaling according to the Bose-glass model is automatically achieved [43,57] using the same $T_g$ and the critical exponents:

$$\nu_{BG} = (D-1)\nu_{VG}/3 \qquad (3)$$

$$z_{BG} = (3z_{VG} - D + 4)/(D-1) \qquad (4)$$

However, in order to be physically acceptable, the critical exponents must lie within a given range of values. These are predicted by the theory [9,45], and have been consistently found in experiments. For a vortex-glass, these are $1 \lesssim \nu_{VG} \lesssim 2$; $4 \lesssim z_{VG} \lesssim 6$ [43,44,57,58]. For a Bose-glass $0.8 \lesssim \nu_{BG} \lesssim 1.8$; $6 \lesssim z_{BG} \lesssim 9$ [44,57,59]. As we show below, checking whether the critical exponents lie within the range of acceptable values will allow us to discriminate between Bose and vortex-glass.

Prior to the data scaling, we independently estimated $T_g$ by studying the derivative of the *V-I* curves. In particular, $d[\log(V)]/d[\log(I)]$ vs. *I* was numerically calculated from the raw data. An example is shown in the inset of figure 2(b). By directly inspecting those derivatives, one can discriminate [58] the isotherms clearly above $T_g$ (whose slope decreases in the low-current limit, reddish colour), from those below $T_g$ (their slope increases with decreasing current within the entire experimental window, bluish colour). Accordingly, one can establish an upper limit for $T_g$ at T=30K and a lower one at T=28K [see horizontal lines in the inset of figure 2(b)]. When looking for the parameters *D*, $T_g$, $\nu_{VG}$ and $z_{VG}$ to collapse the *V-I* sets, $T_g$ was always kept within the range determined by this analysis.

Successful data collapses −as the one shown in figure 2(b)− were obtained for every one of the *V-I* sets available, each corresponding to a different applied magnetic field. Table I summarises the magnetic fields for which the sets of *V-I* were measured, in which *θ* is the angle between the applied field and the YBCO c-axis. For each field, the set of scaling parameters $T_g$, *D*, $\nu$ and *z* that yield the best data collapse is indicated. Irrespectively of the applied magnetic field magnitude and direction, only the vortex-glass scaling with D=2 yielded acceptable critical exponents. For a D=3 vortex-glass, unphysical $z_{VG}$~10.1 are required to achieve the data collapse. For the Bose-glass model as well, only unphysical $z_{BG}$~15.6 allow the data collapse. We conclude from this that a 3D vortex-glass as well as a Bose-glass are to be ruled out.

To independently verify the consistency of the scaling parameters, we used again the derivative of the *V-I* sets. As explained above, this allows us to identify the *V-I* isotherms around $T_g$. The vortex-glass theory predicts [45] that the critical isotherm (i.e. at $T_g$) fulfils $V \propto I^{\alpha+1}$, with $\alpha = (z_{VG} + 2 - D)/(D - 1)$. Thus, for the data displayed in the inset of figure 2(b) we can estimate the range of values for the critical exponent $5 \lesssim z_{VG} \lesssim 5.8$ for D=2. This analysis was performed for every one of the *V-I* sets listed in Table 1. In all cases, the values of $z_{VG}$ obtained from the data collapses with D=2 are consistent with those expected from the derivatives' analysis.

What we learn from the collection of measurements in Table I is the following: i) the critical scaling parameters are essentially the same regardless of the magnitude and direction of the applied field −and in particular, regardless of whether the measurements are done at a matching field or not; ii) in all cases, a three dimensional scaling and a Bose-glass are to be ruled out −instead, a quasi-two-dimensional (D=2) behaviour is observed; and iii) as shown in the inset of figure 2(a), $T_g(B)$ is non-monotonic, $T_g$ being enhanced at the matching fields. While the latter is somewhat expected −if one considers the earlier experiments on $Bi_2Sr_2CaCu_2O_8$ summarised in the introduction [18]− the observations i) and ii) are not. These imply that the glass-transition dimensionality is reduced from D=3 (observed in pristine YBCO films [44,45,57]) to D=2 due to masked ion irradiation. Furthermore, this is independent of whether the flux lattice matches or not the periodic pinning array. The implications of these findings will be unfolded further below, in the discussion section.

3.2 *Angular dependence of the magneto-resistance.*

Figure 3(a) shows the mixed-state magneto-resistance for different directions ($\theta$) of the applied field. The measurements were done in constant Lorentz force geometry (the electrical current flows always perpendicular to the applied field). One can see that the matching fields gradually shift to higher values as the field is rotated off the YBCO c-axis. In figure 3(b), a set of curves as those shown in figure 3(a) is displayed, but in this case the x-axis has been scaled by multiplying $B$ by $\cos(\theta)$. That is, figure 3(b) shows the mixed-state resistance as a function of the component of the applied field parallel to the c-axis (perpendicular to the film plane). This representation shows two things. First, the position of the magneto-resistance minima is dictated solely by the perpendicular component of the applied field, i.e. the matching condition is $B\cos(\theta) = \pm nB_\phi$. This is also shown in the inset of figure 3(b), which displays the matching field corresponding to $n=1$ for different angles $\theta$ as a function of $1/\cos(\theta)$. A linear correlation of slope 1 is observed. The different symbols correspond to two different temperatures, one above and the other below $T_g$ in the relevant field range ($B\cos(\theta) \lesssim 2B_\phi$). That is, this behaviour is observed regardless of whether the system is in the glass phase or not. The second thing we learn from figure 3(b) is that the system has a finite anisotropy: note that, although the matching effects are solely dictated by the component of the applied field perpendicular to the film plane, the background magneto-resistance does depend on the in-plane component $B\sin(\theta)$. Otherwise, a perfect collapse of all of the $R(B)$ curves would be observed in figure 3(b).

To gain further insight into the angular dependence of the magneto-resistance, we performed a series of $R(\theta)$ measurements at constant temperature, applied field and Lorentz force. Some examples of those are shown in figure 4(a)-(e) (hollow symbols), for which the temperature was chosen so that T>$T_g$ within the explored field range. Qualitatively, the same behaviour is observed for T<$T_g$ (not shown). One can see that the curves display resistance minima for $\theta = 90°$, i.e. when the field is applied parallel to the *ab* plane, as it is usual in plain YBCO due to its intrinsic anisotropy [60,61]. Superposed to this, one can observe resistance minima due to the different orders *n* of commensurability between the flux lattice and the periodic pinning potential, given by $B\cos(\theta) = \pm nB_\phi$. Figures 4(a) and (b) display the measurements for applied fields $B = 0.5B_\phi$ and $B = B_\phi$, which respectively show the minima corresponding to *n*=0.5 and *n*=1 centred around $\theta = 0°$. Remarkably, for $B = B_\phi$ [figure 4(b)] the minimum corresponding to *n*=1 is as deep as the one at $\theta = \pm 90°$ .. Note that the minima corresponding to *n*=0.5 can be observed also in figure 4(b) around $\theta = \pm 60°$. The curves in figure 4(c) are paradigmatic to illustrate the critical character of the matching effects: the applied field is only slightly above $B_\phi$ (in particular, only 2% and 10% above $B_\phi$ respectively for each of the curves), yet the minimum corresponding to *n*=1 splits into two minima that strictly satisfy the condition $B\cos(\theta) = B_\phi$. This very unusual behaviour is in contrast with that typically seen in the presence of correlated defects and Bose glass behaviour. For the latter, the resistance minimum in $R(\theta)$ due to vortex accommodation within the defects appears for a unique, well defined angle, regardless of the applied field magnitude and, in particular, of whether the field matching condition is satisfied or not [8,48-50]. Figures 4(d) and (e) show the curves for applied fields $B \sim 2B_\phi$ and $B = 3B_\phi$, respectively. One can see that various minima appear in each of the curves, which correspond (besides the minimum at $\theta = \pm 90°$ usual in plain YBCO) to the different *n* orders in the series $B\cos(\theta) = nB_\phi$.

In order to quantitatively analyse the measurements shown in figure 4(a)-(e), we have developed a model to fit the angular magneto-resistance. Provided that the measurements are performed in the thermally activated flux flow regime observed in the low-current limit above $T_g$, we use the expression for the resistance [41]:

$$R(B,T,\theta) = R_0 \cdot exp\left\{\frac{-U(B,T,\theta)}{K_B T}\right\} \qquad (5)$$

where the angular, field and temperature dependent vortex activation energy is [41]

$$U(B,T,\theta) = \frac{\beta}{[B\cdot\varepsilon(\theta)]^\alpha}\left(1 - \frac{T}{T_C}\right) \qquad (6)$$

In this expression, the field dependence of the activation energy is given by $\alpha$, $\beta$ is an energy scale, and the anisotropic behaviour is taken into account using $\varepsilon(\theta) = \sqrt{\cos^2\theta + \varepsilon^2 \sin^2\theta}$, where $\varepsilon$ is a constant. Note that $\varepsilon(\theta)$ comes from the scaling approach developed by Blatter *et al.* [62], which allows one to describe the angular dependent magneto-resistance of anisotropic superconductors in terms of an effective field $B_{eff} = B\varepsilon(\theta)$ –and in the limit $\varepsilon \to 0$ becomes the two-dimensional Kes' model applied to the highly anisotropic superconductors [63]. In the present experiments, however, such a scaling is not possible [see Fig. 3 (b)]. That is, there is no single parameter ε which is valid within the entire angular range. Nevertheless, as we explain below, Eq. 6 can be used to model the measurements in Fig. 4 if one separately considers the various sources of pinning present in the studied samples, and assigns a different ε to each of them.

To build our model, we assumed that each of the sources of pinning (intrinsic and artificial) has its own characteristic angular-dependent activation energy $U_i$. Regarding the periodic pinning effects at $B\cos(\theta) = nB_\phi$, the model deals with each matching order *n* individually, virtually as if they originated from a different source of pinning. According to this, we expressed the total vortex activation energy as the sum of the activation energies from each source of pinning, $U = \sum_i U_i$, and we derived from Eq. 6 the following expression to fit the experimental results:

$$R = R_0 \cdot exp\left\{\sum_i \frac{-C_i}{[B\cdot\varepsilon_i(\theta-\delta_i)]^{\alpha_i}}\right\} \qquad (7)$$

Since the measurements are performed at constant temperature, we have compressed the energy scale $\beta$ and temperature coefficients in Eq. 6 into a single fitting parameter $C_i$. $\delta_i$ is related to the angle at which the pinning from a source *i* is enhanced; in particular, $90° - \delta_i$ is the angle at which the activation energy $U_i$ is maximum. Note that $\delta_i$ is not a fitting parameter. Contrarily, it has a well-defined value for each of the sources of pinning. For the non-periodic pinning, which is stronger when the field is parallel to the *ab* plane, $\delta_{ab} = 0$. For each of the orders *n* of the matching effects, $\delta_n = 90° - arcos(n \cdot B_\phi/B)$, where *n* takes the values 0.5, 1, 2 and 3. We chose $\alpha_i = 1$ in all cases, based on previous results obtained on YBCO thin film samples [61]. In conclusion, we are left with $R_0$ and a series of pairs $C_i$, $\varepsilon_i$ as fitting parameters. Note that the latter bear the relevant physical meaning: $C_i$ tells us about the pinning strength or pinning energy, and $\varepsilon_i$ about the effective anisotropy of each particular source of pinning.

We used Eq. 7 to fit the experimental data. The agreement between the experimental and the fitted curves is remarkable, as it can be seen in figures 4(a), (b), (d) and (e), in which the solid line on top of the hollow symbols corresponds to the best fit using the model.

Figure 4(f) displays the parameters $C_i$, $\varepsilon_i$ obtained from the curves fit, as a function of the applied magnetic field. We found that the parameters describing the non-periodic pinning –$C_{ab}$ and $\varepsilon_{ab}$– are essentially constant as a function of the applied field [hollow circles in figure 4(f)]. Note that the corresponding anisotropy parameter $\gamma_{ab} = 1/\varepsilon_{ab}$~15-20 is much higher here than in pristine YBCO (γ~5-7) [61]. The parameters describing the matching effects strongly depend on the applied magnetic field. Regarding $C_n$ [top panel in figure 4(f)], we can see that $C_1$ is the largest at any field. This is as expected, given that the magneto-resistance minima for *n*=1 are the deepest (see figure 1). It is remarkable that $C_1 \geq C_{ab}$, and specially that $C_1 \gg C_{ab}$ at the matching field $B = B_\phi$. That is, the periodic pinning array is the strongest source of pinning. Otherwise, $C_n$ decreases with increasing field for all $n$. Regarding $\varepsilon_n$ (see figure 4(f), lower panel), in all cases it decreases with increasing applied field, being the highest for each $n$ at the matching field $B = nB_\phi$, for which $\varepsilon_n \sim 0.18 - 0.35$. The origin of the field-dependence of $C_n$ and $\varepsilon_n$ will be explained below, in the discussion section.

## 4. Discussion

We first discuss the implications of the scaling analysis in section 3.1. We showed that the scaling of the *V-I* characteristic was not possible with a D=3 vortex-glass nor a Bose-glass model, but only according to a D=2 vortex-glass model. The latter is often referred to as a *quasi*-two-dimensional scaling [44]. In contrast to the *pure* two-dimensional vortex-glass, in which $T_g$=0 and the vortex-glass correlations exist only along two dimensions [44,45,64], the *quasi*-two-dimensional scaling observed here shows a finite $T_g$ (see Table 1) and implies that vortex-glass correlations exist along the three spatial dimensions [45]. In particular, for D=2 the glass correlations diverge only for two spatial dimensions and remain finite (shorter than the sample's size) along the third one. Considering that we observed D=2 irrespective of whether the flux lattice matches or not the periodic pinning array (see table I) −that is, regardless of the constraints imposed within the *ab* plane− it is reasonable to assume that glass correlations diverge in-plane and remain finite along the c-axis. In conclusion, an important effect of the masked O[+] ion irradiation is that it reduces vortex-glass correlations along the c-axis. This is supported by the relatively large anisotropy $\gamma_{ab} \sim$ 15-20 deduced from the angular magneto-resistance fits, which suggests weaker coupling between $CuO_2$ planes than in pristine YBCO [45]. In this respect −exclusively concerning the D=2 scaling and the anisotropy $\gamma_{ab}$− the behaviour observed here resembles that of oxygen-depleted YBCO thin films [44]. This can be understood if one considers that, although ion irradiation induces the largest amount of disorder in the hole areas directly exposed to the O[+] beam by the mask, a significant amount of disorder is induced also in the inter-hole areas not directly exposed to the beam [36]. Since the induced disorder is mainly in the form of oxygen vacancies and interstitials [65], one indeed expects that its effect on the system's dimensionality is similar to that caused by oxygen depletion [44].

The angular dependence of the magneto-resistance also differs from that expected in typical Bose-glass behaviour. In a Bose-glass, a resistance minimum should appear in R(θ) for a well-defined angle, that is, when the magnetic field is applied along a particular direction, independently of the field magnitude [8,48-50]. Contrary to this, in the present experiments the resistance minima in R(θ) may appear for any direction of the applied field 0<θ≤90º, depending of the field magnitude. In a Bose-glass, vortex localization within the pinning defects occurs when the vortex-lines are parallel to a particular direction. Here, vortex localization occurs for particular values of the out-of-plane component of the magnetic field, that is, for certain in-plane vortex densities. This implies that vortex localization in the artificial pinning sites is dictated here by the vortex-vortex interactions within the *ab* plane −at variance to Bose-glass systems, in which the vortex line tension determines the angle of accommodation within the defects [9,66]. In conclusion, in the present system in-plane correlations between vortices dominate over correlations along the vortex line.

We discuss now the pictures of the vortex accommodation to the pinning landscape in tilted applied fields. One possibility is to consider that vortices go straight across the artificial pinning sites, as sketched in figure 5(a). A second possibility is shown in figures (b) or (c): vortices would present a kinked structure consisting of stacks of correlated pancake vortices linked by Josephson strings that lie in the *ab*-plane [67]. The second possibility seems more likely, provided the relatively high-anisotropy $\gamma_{ab}$ and the limited c-axis correlations implied by the scaling analysis. In fact, recent experiments support this type of kinked vortex structure in underdoped YBCO [68]. Note that the Josephson strings are necessary to conserve the in-plane component of the applied field, and consequently their number must increase with the angle as sketched in figure 5(b)-(c). For any of the two possibilities shown in figure 5(a) and (b)-(c), the vortex line fraction pinned within the irradiated regions decreases with increasing angle. One expects therefore that the pinning energy due to the commensurability between the flux-lattice and the periodic array decreases with increasing tilt angles. This allows for an understanding of the decrease of $C_n$ [figure 4(f), top panel] as $B$ is increased and the matching condition $B\cos(\theta) = nB_\phi$ is satisfied at angles closer to $\theta = 90°$ [see e.g. figures 4(d) and (e)]. Note that $\varepsilon_n$ also decreases with increasing applied magnetic field [figure 4(f), bottom panel], which accounts for the observation that the $n$ resistance minima in $R(\theta)$ become much narrower as they shift towards $\theta = \pm 90°$. At first sight, this may suggest that the matching of the flux-lattice to the periodic pinning array becomes more critical as it is achieved in increasingly tilted fields. However, this conclusion should not be pushed too far, since that behaviour can otherwise be understood by considering that, as the minima shift towards $\theta = 90°$ in increasing fields $B \gg B_\phi$, the derivative of the field component along the c-axis $d[B\cos(\theta)]/d\theta$ increases. That is, the closer to $\theta = 90°$ a particular minimum sits, the faster we go "in" and "out" of the matching condition when the field is rotated, and the narrower the resistance minima. Within the model of equation (7), this requires that $\varepsilon_n$ decreases as $B$ increases and the $n$ minimum shifts towards $\theta = 90°$.

Before concluding, we draw some comparisons between the present experiments and earlier ones in which different materials and pinning centres were used. It is particularly interesting that a Bose-glass behaviour is not found here. The key characteristics of our system are i) the relatively high anisotropy as compared to pristine YBCO, ii) the periodic ordering of the artificial pinning sites, and iii) the presence of strong *random* pinning that coexists with the periodic one. None of these characteristics can explain the observed behaviour when considered separately: i) a Bose-glass has indeed been observed in more anisotropic superconductors −e.g. $Tl_2Ba_2CaCu_2O_8$ [11] and $Bi_2Sr_2CaCu_2O_8$ [66]− with columnar defects, ii) a Mott insulator phase (a particular case of the Bose-glass) has been previously observed in $Bi_2Sr_2CaCu_2O_8$ single crystals with periodic arrays of holes [18], and finally, iii) a crossover from vortex-glass into Bose glass has been observed in YBCO thin films in which strong random pinning coexists with c-axis correlated pinning ($BaZrO_3$ nanorods) [57]. From the above, we conclude that it is the concurrence of strong quenched disorder and relatively high anisotropy that impedes the observation of Bose-glass behaviour in our samples, even in presence of a periodic pinning array. We anticipate that if periodic pinning sites created by irradiation could be obtained without increasing the YBCO anisotropy, a crossover from vortex to Bose-glass behaviour should be observed.

## 5. Conclusions

The presence of a nanoscale periodic pinning array induced via masked $O^+$ ion irradiation in YBCO thin films strongly modifies the glass transition behaviour and the angular dependent magneto-resistance in the mixed-state. The 3D vortex-glass transition typically observed in these films becomes a *quasi*-two-dimensional glass transition after irradiation, pointing to a weakening of the vortex correlations along the c-axis. The glass transition temperature is enhanced at the matching fields [inset figure 2(a)], and the magneto-resistance shows pronounced minima which are solely governed by the applied field component parallel to the c-axis [figure 3(a)]. Thus, vortex localization within the artificial pinning defects does not occur for particular directions of the applied field, but for particular in-plane vortex densities. This is in contrast to the systems in which correlated defects stabilise a Bose-glass phase. We conclude that the concurrence of enhanced anisotropy and random pinning preclude the observation of a Bose-glass phase even in the presence of strong periodic pinning.

**Acknowledgements:**

We acknowledge Prof. Z. Sefroui for discussions. Work supported by French ANR « SUPERHYBRIDS-II » and « MASTHER » grants, and by MICI (MAT2011-28874-C02-01 and Consolider Nanoselect). J. T. acknowledges the support of Fundación Barrié (Galicia, Spain) and V. R. from the JAE-CSIC PhD grant.

**Figure Captions**

**Figure 1:** Resistance as a function of the magnetic field (applied parallel to the c-axis), measured at T=0.73$T_{c0}$ with an injected current density J=25 kA cm$^{-2}$. $T_{c0}$=48 K is the temperature at which R($T_{c0}$)=10$^{-2}$$R_N$ in zero field, with $R_N$ the normal-state resistance at the onset of the transition. The order of matching *n* is indicated by the labels. Inset: Scanning electron microscopy image of a PMMA e-beam lithographed mask used for ion irradiation.

**Figure 2:** (a) Set of voltage versus current isotherms in a field $B = 2B_\phi$ applied parallel to the c-axis, for temperatures ranging from T=50 K (bright red) to T=20 K (bright blue). Isotherms are ~1 K apart. Inset: Glass transitions temperature $T_g$ versus magnetic field (applied parallel to the c-axis), as obtained from the I-V collapses. The line is a guide to the eye. (b) Collapse of the *V-I* curves in (a) according to the vortex-glass scaling discussed in the main text. The scaling parameters are shown in Table 1. Inset: Derivative $d(\log V)/d(\log I)$ of the *V-I* set in (a). The horizontal dashed lines delimit the region within which the critical isotherm must lie.

**Figure 3:** (a) Resistance as a function of the magnetic field (normalised to $B_\phi$) for different field directions $\theta$ indicated by the label. Measured at T=0.87 $T_{c0}$ and with J=2.5 kA cm$^{-2}$. The first matching field observed for each angle is marked with an arrow. Inset: Scheme with the definition of $\theta$. (b) Resistance as a function of the magnetic field component parallel to the c-axis $B cos(\theta)$, normalised to $B_\phi$, for different field directions $\theta$=0, 30, 40, 45, 65 and 70 degrees (bottom to top). Measured at T=0.87 $T_{c0}$ and with J=2.5 kA cm$^{-2}$. Inset: matching field corresponding to *n*=1 as a function of 1/cos($\theta$), for T=49 K≈$T_{c0}$ (triangles) and T=0.52 $T_{c0}$ (circles).

**Figure 4:** (a)-(e) Resistance versus applied field direction $\theta$ with constant field magnitude (indicated by the labels in factors of $B_\phi$), measured at T=40 K with an injected current J=2.5 kA cm$^{-2}$ (hollow circles). The solid lines are the best fit using the model described in the text. The labels indicate the matching order *n* associated to the minima. (f), (g) Fitting parameters $C_n$ and $\varepsilon_n$ as a function of the applied magnetic field normalised to $B_\phi$.

**Figure 5:** Upper panel: Top view of the magnetic flux distribution in the *n*=1 matching condition. Bottom: Cross section of the same state. Two different possible vortex accommodations are shown: (a) Straight vortex lines; (b),(c) kinked stacks of pancake vortices for two different tilt angles. Vortex lines are represented by (red/orange, for clarity) lines with arrowheads that show the magnetic field direction. The strongly irradiated regions that behave as pinning sites are shown with darker (green) filling.

**Table I:** Scaling parameters for *V-I* sets (collection of isotherms measured in a fixed applied field B and angle θ at different temperatures) according to the vortex-glass model. D is the dimensionality, $\upsilon$ and z the static and dynamic critical exponents, $z_{slope}$ the dynamic critical exponent expected from the slope of the I-V curves nearby the critical isotherm, and $T_g$ the glass transition temperature.

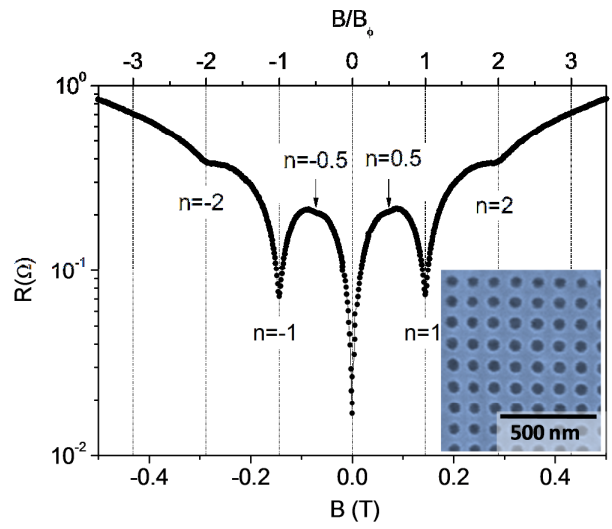

**Figure 1**

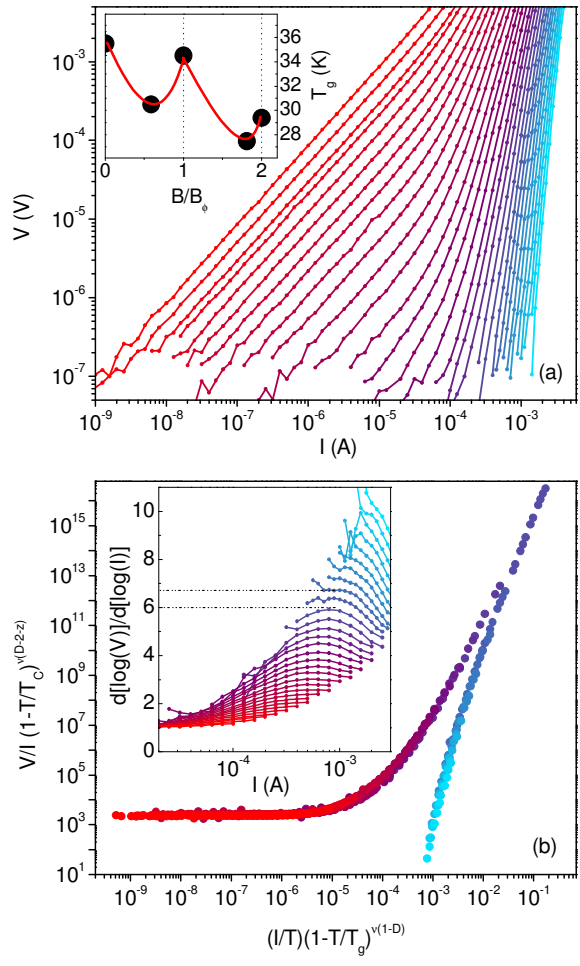

**Figure 2**

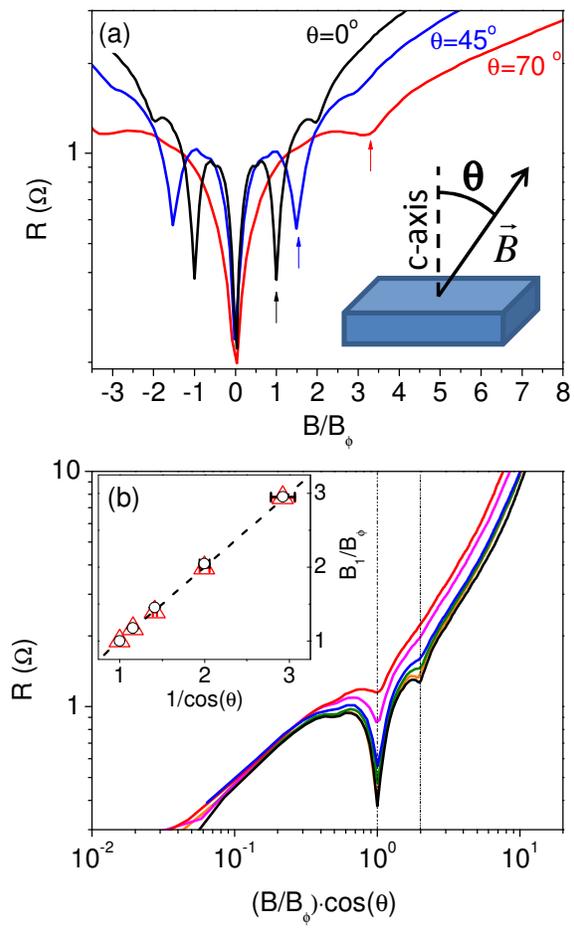

**Figure 3**

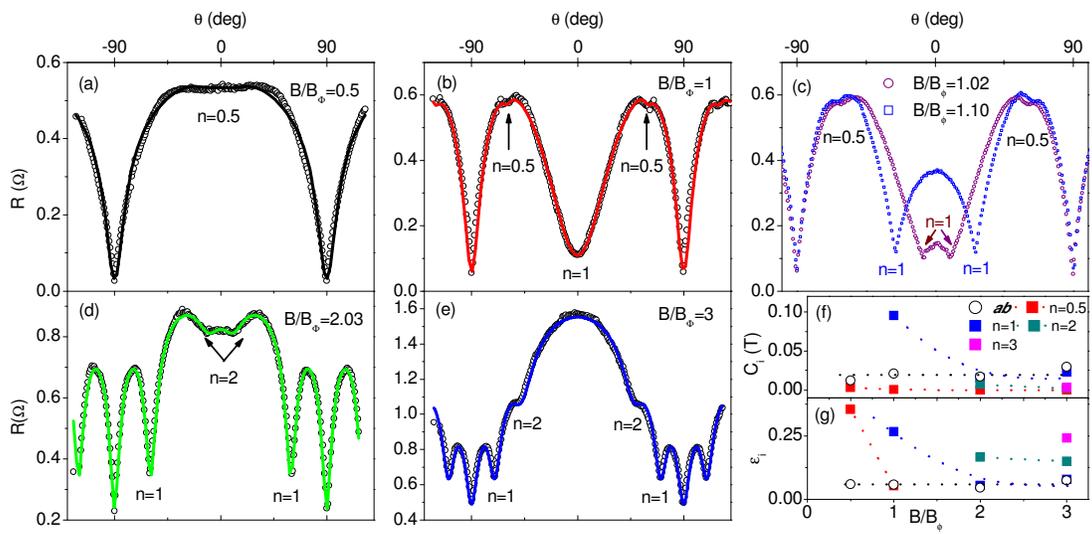

**Figure 4**

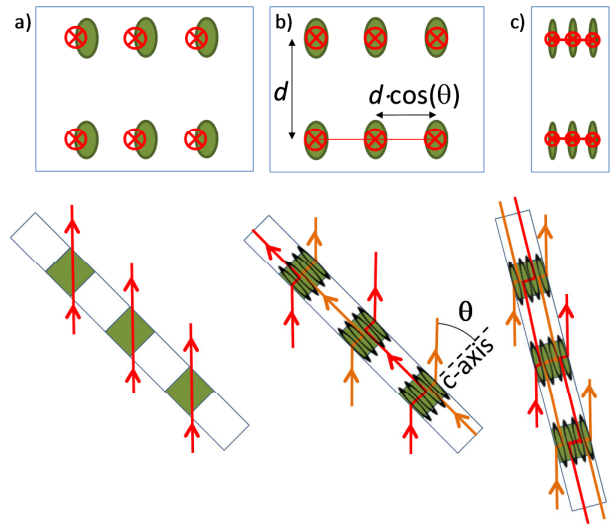

**Figure 5**

| B (T) | θ (deg) | B·cos(θ)/B$_\varphi$ | D | $\upsilon \pm$ ~0.1 | $z \pm$ ~0.2 | $z_{slope}$ | $T_g \pm 0.5$ (K) |
|---|---|---|---|---|---|---|---|
| 0.085 | 0 | 0.59 | 2 | 2 | 4.8 | 4.9 – 5.4 | 30.5 |
| 0.143 | 0 | 1 | 2 | 2 | 5 | 5.4 – 5.7 | 34.5 |
| 0.260 | 0 | 1.81 | 2 | 1.9 | 5.5 | 5.6 – 6.1 | 27.5 |
| 0.280 | 0 | 2 | 2 | 1.8 | 5 | 5 – 5.8 | 29.4 |
| 0.160 | 27 | 1 | 2 | 2 | 4.8 | 5–6 | 34.5 |
| 0.250 | 55 | 1 | 2 | 2 | 5.2 | 5-5.8 | 32.5 |
| 0.565 | 90 | - | 2 | 1.6 | 4.5 | 5-5.8 | 30.5 |

**Table I**